\documentclass[aps,rmp,final,nolinenumbers,amsmath,amssymb,graphicx,longbibliography,a4paper,11pt]{revtex4-1}

\pdfoutput=1

\usepackage[version=4]{mhchem}
\usepackage[separate-uncertainty,detect-all,range-phrase=\ and\ ]{siunitx}
\DeclareSIUnit\ppm{ppm}
\DeclareSIUnit\day{d}
\DeclareSIUnit\year{yr}
\DeclareSIUnit\ton{t}
\DeclareSIUnit[number-unit-product = {\,}]\xu{xu}
\usepackage[utf8]{inputenc}
\usepackage[T1]{fontenc}
\usepackage{bm}
\usepackage[english]{isodate}
\cleanlookdateon
\usepackage{hyperref}
\begin{document}

\title{Who discovered positron annihilation?}
\date{\printdate{31.12.2021}}
\author{Tim Dunker}

\begin{abstract}
In the early 1930s, the positron, pair production, and, at last, positron annihilation were discovered. Over the years, several scientists---commonly, Thibaud and Joliot---have been credited with the discovery of the annihilation radiation. A conversation between Werner Heisenberg and Theodor Heiting prompted me to examine relevant publications, when these were submitted and published, and how experimental results were interpreted in the relevant articles. I argue that it was Theodor Heiting---usually not mentioned at all in relevant publications---who discovered positron annihilation, and that he should receive proper credit.

\end{abstract}

\maketitle

\section{Introduction}
\label{sec:intro}
There is no doubt that the positron, the electron's anti--particle, was discovered by Carl D. Anderson \citep[e.g.][]{Anderson1932,Hanson1961,LeoneRobotti2012} after its theoretical prediction by Paul A. M. Dirac \citep{Dirac1928,Dirac1931}. Further, it is undoubted that Patrick M. S. Blackett and Giovanni P. S. Occhialini discovered pair production by taking photographs of electrons and positrons created from cosmic rays in a Wilson cloud chamber \citep{BlackettOcchialini1933}. The answer to the question who experimentally discovered the reverse process---positron annihilation---has been less clear. Usually, Fr\'ed\'eric Joliot and  Jean Thibaud receive credit for its discovery \citep[e.g.,][p. 110]{Roque1997}. Several of their contemporaries were enganged in similar research. In a letter correspondence with Werner Heisenberg \citep{HeitingHeisenberg1952}, Theodor Heiting\footnote{See Appendix \ref{ap:bio} for a rudimentary biography.} claimed that it was he who discovered positron annihilation. 

To assess whether this claim has any merit, and find out who discovered positron annihilation, I investigate the literature on positron annihilation from the early 1930s. To be able to answer who should be credited for the discovery, I also evaluate the interpretation of the experimental results in the publications.

\section{The positron, pair production, and positron annihilation}
\label{sec:positron}
Before giving an overview of the relevant publications and assessing the physical interpretations, I briefly sketch a few concepts relevant to positron annihilation.

\subsection{Theoretical prediction and experimental discovery of the positron}
\label{sec:prediction}
In his work ``The quantum theory of the electron'', \citeauthor{Dirac1928} postulated the existence of vacant negative energy states regarding free electrons \citep[][Eq. (9)]{Dirac1928}. In a slightly altered form, that equation reads:
\begin{equation}
 \dfrac{E^2}{c^2}-p_r^2-(mc)^2 = 0,
\end{equation}
where $p_r=-ih\frac{\partial}{\partial x_r}$ and $E=ih\frac{\partial}{\partial t}$. It has become known as the ``Dirac equation''. As \citeauthor{Dirac1928} outlined in his speech when receiving the Nobel Prize in Physics in 1933 \citep{Dirac1933}, this equation can be satisfied either by particles of energy $E>mc^2$, or by particles of energy $E<-mc^2$. Such states of negative energy for the electrons correspond to particles with a positive elementary charge, the positive electron, or positron. \citeauthor{Dirac1928} interpreted unoccupied negative energy states as ``holes'', which were short of negative energy, and therefore had positive energy. His theory \citep[for a review, see e.g.][especially \S\ 23]{HillLandshoff1938} is therefore often called ``hole theory''. These ``holes'' could then be filled by ordinary electrons from higher shells, releasing energy as radiation---that is, positron annihilation. \citeauthor{Dirac1928} also mentioned the opposite process as a possibility \citep{Dirac1928}: transforming electromagnetic radiation into an electron and a positron, that is, pair production\footnote{In fact, pair production is not strictly the opposite process, because in pair production, one photon is transformed into two particles, while positron annihilation transforms two particles into at least two photons. One--photon annihilation is practically irrelevant, as \citet{FermiUhlenbeck1934} calculated.}. Dirac was well aware that protons were far too heavy to be in agreement with his theory \citep{Dirac1930}, but these were at the time the only known positively charged particles.

Not very long after Dirac's postulation, \citet{Anderson1932} reported on the existence of ``positives'', which were ``comparable in mass and magnitude of charge'' to electrons. His discovery confirmed \citeauthor{Dirac1928}'s theoretical postulation, and was based on observations of tracks of positively charged electrons in a Wilson cloud chamber \citep{Anderson1932}. Protons could be ruled out from mass and energy considerations \citep{Anderson1933a}. The new positively charged particle came to be known as a positron.\footnote{According to Carl Anderson, the name ``positron'' was not suggested by him, but by Watson Davis, the then--editor of the journal Science \citep{AndersonWeiner1966}.}

Two processes in which a positron is involved, are pair production and positron annihilation. These are fascinating example of the transformation of light into matter and vice versa, respectively.

\subsection{Pair production}
\label{sec:production}
In pair production, electromagnetic radiation is transformed into matter, namely an electron--positron pair. As \citet{OppenheimerPlesset1933} have shown, if a photon has an energy of at least twice the rest mass of an electron $E=\SI{1.022}{\mega\electronvolt}$, pair production becomes possible:
\begin{equation}
 \gamma \rightarrow e^-\ +\ e^+
\end{equation}
Twice the rest mass of an electron is needed, because a positron has the same rest mass as an electron. The pair production cross--section is 
\begin{equation}
 \sigma_{pp} \approx \alpha Z^2 r^2.
\end{equation}
Here, $Z$ is the atomic number of the element in which pair production takes place, $r=e^2 \left(\num{4}\pi \varepsilon_0 m c^2\right)^{-1}$ is the electron radius, and $\alpha\approx137^{-1}$ is the fine--structure constant.

The creation of electron and positron pairs was first observed by \citet{BlackettOcchialini1933} in a Wilson cloud chamber. Shortly after, \citet{Chadwicketal1934} argued that the positrons are most likely not created in the nucleus itself, but in the ``electric field outside the nucleus'' (that is, in the electron shells).

\subsection{Positron annihilation}
\label{sec:annihilation}
The reverse process of pair production is called positron annihilation. This process has also been predicted by \citeauthor{Dirac1928}'s hole theory \citep{Dirac1928}, although the positron had not been discovered yet, thus \citeauthor{Dirac1930} spoke of ``protons'' and acknowledged the problem of a large difference in mass between an electron and a proton \citep[][p. 361 yo 362]{Dirac1930}. When a positron combines with an electron, two $\gamma$--rays are emitted:
\begin{equation}
e^-\ +\ e^+\ \rightarrow\ \gamma\ +\ \gamma
\end{equation}
Although the probability of annihilation was deemed low \citep{FermiUhlenbeck1934} from a theoretical point of view, this process of annihilation does in fact occur. Before it occurs, a positron and an electron may form a positronium atom: This is a hydrogen--like atom, but with a positron instead of a proton \citep{Deutsch1951}. The singlet state of the positronium atom, so--called parapositronium, decays by emitting two photons of equal energy\footnote{The probability of the emission of a higher (even) number of photons  is not zero, but much smaller than that of the two--photon decay.}. Parapositronium has a lifetime of $\sim\SI{142e-9}{\second}$ in vacuum \citep{Al-RamadhanGidley1994}. 

The emitted $\gamma$--rays in the annihilation process have energies of \SI{511}{\kilo\electronvolt} each, corresponding to the electron rest mass. These $\gamma$--rays are emitted in opposite direction to each other to satisfy the conservation of total momentum. These two quanta are also polarized perpendicularly to each other. In many of the early articles, especially in the 1930s, the wavelength that corresponded to the quantum's kinetic energy was commonly expressed in the ``X--unit/xu'', or ``X.E.'' for ``R\"ontgeneinheit'' in German, or ``UX/unit\'{e} x'' in French. \citet{Bearden1965} has determined the wavelength conversion factor $\Lambda$ to be
\begin{equation*}
\Lambda = \SI{1.002076(5)e-13}{\metre\per\xu}.
\end{equation*}
For some of the articles cited below, I use this conversion factor to transform the outdated units to today's SI units.

\section{Who discovered positron annihilation?}
\label{sec:discovery}
In the early 1930s, many different scientists in France, Germany, and Great Britain worked on the interaction of hard $\gamma$--rays and different metal absorbers. Results were, accordingly, published in English, French, and German. Nomenclature was not yet well--developed, because both the underlying theory \citep{Dirac1928} and many experimental effects were quite new and sometimes uncertain, or at least subject to interpretation. It was a very vibrant field and publications arrived at the scientific journals in very short intervals, and, occasionally, different scientists published their results in the same issue of a journal. In the following, I take the Nobel Prize in Physics in 1948 as a starting point. My goal is to answer the question who first discovered the annihilation radiation. Table \ref{tab:publications} gives an overview over relevant publications, when these were submitted, and when these were finally published.

\begin{table}[!t]
\centering
\caption{Relevant publications on the discovery of positron annihilation, sorted chronologically according to the publication date. Articles in \textit{Comptes rendus hebdomadaires des s\'eances de l'Acad\'emie des sciences} and \textit{Journal de Physique et le Radium} were published in French. Articles in \textit{Naturwissenschaften} and \textit{Zeitschrift f\"ur Physik} were published in German. Articles in the other journals were published in English. Information regarding the receiving date of \citet{Joliot1933}, \citet{Thibaud1933d}, and \citet{Thibaud1934a} was obtained from the Archives of l'Academie des Sciences (Isabelle Maurin--Joffre, pers. comm., \printdate{17.02.2020}).}

\label{tab:publications}
\begin{ruledtabular}
\begin{tabular}{llrr}
Author & Journal & Submitted or received & Published\\
\cline{1-4}
\citet{Heiting1933a}       & Naturwissenschaften    & \printdate{07.08.1933} & \printdate{15.09.1933} \\
\citet{Heiting1933b}       & Naturwissenschaften    & \printdate{28.09.1933} & \printdate{10.11.1933} \\
\citet{Joliot1933}         & C. R. Hebd. Acad. Sci. & rec. \num{15} days before publication                & \printdate{18.12.1933} \\
\citet{Thibaud1933d}       & C. R. Hebd. Acad. Sci. & rec. \num{15} days before publication               & \printdate{18.12.1933} \\
\citet{Heiting1934}        & Z. Phys.               & \printdate{02.11.1933} & January 1934 \\
\citet{GrayTarrant1934a}   & Proc. R. Soc. Lond. A  & \printdate{07.11.1933} & \printdate{01.02.1934} \\
\citet{Thibaud1934a}       & C. R. Hebd. Acad. Sci. & rec. \num{15} days before publication                & \printdate{05.02.1934} \\
\citet{BotheHorn1934a}     & Naturwissenschaften    & \printdate{10.01.1934} & \printdate{16.02.1934} \\
\citet{Gentner1934a}       & J. Phys. Radium        & \printdate{01.01.1934} & February 1934 \\
\citet{Williams1934}       & Nature                 & \printdate{22.01.1934} & \printdate{17.03.1934} \\
\citet{Gentner1934b}       & Naturwissenschaften    & \printdate{15.05.1934} & \printdate{22.06.1934} \\
\citet{Thibaud1934b}       & Phys. Rev.             & \printdate{12.04.1934} & \printdate{01.06.1934} \\
\citet{Joliot1934b}        & J. Phys. Radium        & \printdate{23.05.1934} & July 1934 \\
\citet{Klemperer1934}      & Math. Proc. Cambridge  & \printdate{20.05.1934} & July 1934\\
\citet{BotheHorn1934b}     & Z. Phys.               & \printdate{06.03.1934} & September 1934 \\
\citet{CraneLauritsen1934} & Phys. Rev.             & \printdate{01.03.1934} & June 1934 \\

\end{tabular}
\end{ruledtabular}
\end{table}

In his presentation speech of the Nobel Prize in Physics in 1948, which was awarded to P. M. S. Blackett, G. Ising of the Nobel Committee for Physics said the following \citep[][p. 95]{Ising1964}:

\begin{quote}
``[...] Reversely, the meeting of two slow electrons, opposite in sign, results in their fusion and annihilation as material particles; in this process two light quanta, each of 1/2 million electron volts, are formed; these fly out from the point of encounter in opposite directions, so that the total momentum remains about zero (for even light possesses a momentum directed along the ray). Blackett and Occhialini immediately drew these conclusions from their experiments and were guided in so doing by the earlier mathematical electron theory elaborated by Dirac on the quantum basis. The existence of the `annihilation radiation' was shortly afterwards established experimentally by Thibaud and Joliot. [...]'',
\end{quote}
referring to the works of \citet{Thibaud1933d} and \citet{Joliot1933}.

In December \citeyear{Joliot1933}, \citeauthor{Joliot1933} published his results on positron annihilation \citep{Joliot1933}. Five months later, he submitted a more detailed version of those experimental results, which were published in July 1934 \citep{Joliot1934b}.

\citet[][p. 1631, ``6.'']{Thibaud1933d} claimed to have published the first experimental proof of positron annihilation radiation. Shortly after, \citet{Gentner1934b} mentioned that the experimental proof for positron annihilation, i.e. the detection of a $\sim \SI{500}{\kilo\electronvolt}$ radiation component, was achieved simultaneously by \citet{Joliot1933} and \citet{Thibaud1933d}. \citet{Joliot1933} and \citet{Thibaud1933d} published their results in the same issue of Comptes Rendus, dated \printdate{18.12.1933}. 

\citet{Thibaud1934b} subsequently published a more detailed version in English. He comments on \citeauthor{Joliot1933}'s results \citep{Joliot1933}, stating that \SI{86}{\percent} of \citeauthor{Joliot1933}'s results obtained with a Geiger--M\"uller counter were due to ``parasitic effects'', while his own method (film exposure) was exempt from such contamination \citep[see also][]{Thibaud1934a}.

Results similar to \citeauthor{Joliot1933} and \citeauthor{Thibaud1933d} have been obtained by \citet{GrayTarrant1934a}, namely, that much of the secondary $\gamma$--radiation occurred with energies $\SI{0.5e6}{\electronvolt} \leq E_{kin} \leq \SI{1.0e6}{\electronvolt}$, regardless of the scattering angles observed. They used various absorber materials \citep{GrayTarrant1934a}. They interpreted their results in a companion article \citep{GrayTarrant1934b}, but did not relate their results to \citeauthor{Dirac1928}'s theory.

\citet{Williams1934} detected backscattered particles, resulting from hard $\gamma$--rays incident on a \ce{Pb} plate. The backscattered particles were detected by an ionisation chamber at an angle of \ang{140} relative to the \ce{Pb} plate. He performed two different experiments: one without an additional absorber, and one with an \ce{Al} plate as absorber between the \ce{Pb} plate and the ionisation chamber. He interpreted the results such that, when the \ce{Al} plate was present, most positrons emitted by the \ce{Pb} plate were absorbed by the \ce{Al} plate, while the positrons without the absorber would be annihilated in greater distance from their source. However, \citeauthor{Williams1934} could only speculate about the particles being positrons, and the experiments did not yield information on the particles' energy or charge, or energy of the annihilation radiation.

\citet{Klemperer1934} has also provided proof of the annihilation radiation, using a Geiger--M\"uller counter in his experiments. His report was received by the Cambdrige Philosophical Society on 20 May 1934.
 
\citet{Gentner1934b} tried to clarify the differences between theoretical expectations according to the Klein--Nishina formula and the spectral dependence of absorpion coefficients found experimentally. \citeauthor{Gentner1934b} interpreted his own results as Compton electrons, and to lesser extent as electrons from a photoeffect in the absorber which were backscattered towards the filter. He concluded that, because the bremsstrahlung spectrum consists of energies up to \SI{e6}{\electronvolt}, many of the characteristics of secondary radiation might be explained by backscattered electrons \citep{Gentner1934b}. 

In their review article, \citet[][\S 23, pp. 37 to 40]{FleischmannBothe1934} \footnote{I have not been able to determine which article of Thibaud their footnote 177 refers to.} credit \citet{Joliot1934a}, \citet{Thibaud1933d,Thibaud1934a}, and \citet{CraneLauritsen1934} with the experimental discovery of positron annihilation. Heiting's work \citep{Heiting1934} is mentioned, too, in connection with the secondary radiation observations made with hard $\gamma$--rays.  \citet[][p. 67]{Amaldi1984} also credits \citet{Joliot1933,Joliot1934b} and \citet{Thibaud1933d} with the discovery, while \citet[][footnote 56a, p. 121]{HillLandshoff1938} attribute the experimental proof to \citet{Klemperer1934} and \citet{CraneLauritsen1934}. More recently, \citet[][]{Guerraetal2012} indirectly credited Joliot with the discovery of positron annihilation, writing that 
\begin{quote}
``[...] between the middle of December 1933 and early January 1934, [Joliot] studied positron--electron annihilation, finding results in complete agreement with Dirac’s theory, which he published in the \textit{Comptes rendus}'',
\end{quote}
referring to \citet{Joliot1933} and \citet{Joliot1934a}. Evidently, all of the works cited in this paragraph have been published\footnote{For three articles, I have been unable to find the date of submission or reception by the journal ``Comptes rendus''. Judging from the publication date, though, it seems certain that these articles were submitted after Heiting had submitted his two first articles. It seems very likely---given the rapid nature of publication at ``Comptes rendus'' with its two--week cycle between issues---that these were submitted after the publication of Heiting's first two articles.} after Heiting's first two articles (cf. Table \ref{tab:publications}).
 
Let us return to the contents of the Nobel Prize speech: in \citeyear{DPG1951}, Physikalische Bl\"atter\footnote{This was the journal of the Verband Deutscher Physikalischer Gesellschaften, at that time the intermediate organization of the Deutsche Physikalische Gesellschaft.} published articles on recently awarded Nobel Prizes in Physics \citep{DPG1951}. Among them was the Nobel Prize 1948, awarded to P. M. S. Blackett. This prompted a comment by Theodor Heiting and a reply by Werner Heisenberg \citep{HeitingHeisenberg1952}. Heiting claimed that he had discovered positron annihilation even earlier than Joliot and Thibaud. Heisenberg, whom Heiting thanked in his publications for his counselling and persistent interest in his work, replied that he could dimly remember a conversation with Heiting and a Mr Messerschmidt\footnote{This was almost certainly \href{http://www.catalogus-professorum-halensis.de/messerschmidt-wilhelm.html}{Wilhelm Messerschmidt} \citep[see also][]{Bruecheetal1966}, who, like Heiting, was a member of Gerhard Hoffmann's group in Halle in the early 1930s.}, and that he was happy about Heiting's confirmation of Dirac's theory \citep{HeitingHeisenberg1952}\footnote{Apparently, Heiting got a private lecture by Werner Heisenberg on Dirac's ``hole'' theory. I have not found other sources regarding the communication between Heiting and Heisenberg}. Heisenberg remarked that it might have been possible that also other physicists before Heiting mentioned positron annihilation radiation, or made it experimentally probable. However, he could not find any evidence for that, and subsequently Heiting should be named as the discoverer of positron annihilation radiation. 

Heiting was a doctoral student with Gerhard Hoffmann at the Institut f\"ur Experimentalphysik in Halle an der Saale, Germany.\footnote{\label{fn:uni}At that time, the university was still called Vereinigte Friedrichs--Universit{\"a}t Halle--Wittenberg. On \printdate{10.11.1933}, the university got today's name ``Martin--Luther--Universit\"at Halle--Wittenberg''. The university's official name includes Wittenberg, but the actual location is Halle. There has not been any university activity in Wittenberg, approximately \SI{60}{\kilo\metre} from Halle, for a long time (Reinhard Krause--Rehberg, pers. comm., 2018).} 
Three articles were published by \citeauthor{Heiting1933a}, submitted on \printdate{07.08.1933} \citep{Heiting1933a}, \printdate{28.09.1933} \citep{Heiting1933b}, and the third one received by the journal on \printdate{02.11.1933} \citep{Heiting1934}. This is six weeks earlier than \citeauthor{Joliot1933}'s and \citeauthor{Thibaud1933d}'s publications \citep{Joliot1933,Thibaud1933d}.

In his experiments, \citeauthor{Heiting1933a} used $\gamma$--rays from a \ce{RaTh} source, which was equivalent to \SI{28}{\milli\gram} of radium \citep{Heiting1934}. This material emits electromagnetic radiation with a wavelength of \SI{4.7}{\xu} \citep{Heiting1934}, which equals 
\begin{equation*}
\lambda=\SI{4.7}{\xu} \hspace{0.25cm}\Rightarrow \hspace{0.25cm} E_{kin}=\SI{2.63e6}{\electronvolt} \approx \num{5.2}\cdot m_ec^2,
\end{equation*}
corresponding to the radioactive decay \ce{Th}C'', that is, the transition $\ce{^208Tl} \to \ce{^208Pb}$. This kinetic energy is sufficient to create electron--positron pairs.

The $\gamma$--rays were incident on aluminium, iron, copper, and lead absorbers, respectively. An ionization chamber was used to detect the penetrating radiation. \citeauthor{Heiting1933a} measured the attenuation coefficients for all the absorber materials, and reported the first results in August 1933 \citep{Heiting1933a}. \citeauthor{Heiting1933a} discovered that all the mentioned absorbers emitted a type of secondary radiation, which was independent of the absorber element \citep{Heiting1933a,Heiting1933b}.  Using \citeauthor{Chao1930}'s method, who found that the attenuation coefficient is a function of wavelength \citep{Chao1930}, \citeauthor{Heiting1933a} determined the mean wavelength from the measured attenuation coefficients to be
\begin{equation*}
\lambda=\SI{23.8 +- 1.0}{\xu} \cdot \Lambda = \SI{2.385e-12}{\metre}.
\end{equation*}
With today's values of the Planck constant and the speed of light, this wavelength corresponds to a photon energy of $E_{kin}=\SI{0.52(2)e6}{\electronvolt}$. Well within the measurement uncertainty, this energy equals an electron's rest mass. This finding was pointed out by \citeauthor{Heiting1933b} in his second publication on the topic \citep{Heiting1933b}, and he also interpreted that secondary radiation as recombination radiation (``Rekombinationsstrahlung'') of an electron with a positron\footnote{`` The fact that the wavelength $\lambda = \num{24}\, \text{xu}$ occurs, requires special attention, because this wavelength corresponds to the energy $h\nu = m_0c^2$. Recently, with a Wilson cloud chamber, it was found \citep{Anderson1933b} that positrons can occur in heavy elements when these are hit by hard \ce{Th} C''--$\gamma$ rays. According to \textsc{Dirac}'s view, the positive electron only has a short lifetime, because it seeks to recombine with a negative electron, emitting the energy $2m_0c^2$ as two quanta $h\nu = m_0c^2 (\lambda = \num[output-decimal-marker={.}]{24.2}\, \text{xu})$. The observed secondary radiation can thus be interpreted as recombination radiation, and the nuclear absorption can be interpreted as splitting of the primary $\gamma$ rays in the nucleus into positive and negative electrons. [...]''
 
}: positron annihilation radiation.

The two major conclusions, out of a total of four, in his most comprehensive article \citep[][p. 137]{Heiting1934} are:
\begin{quote}
\begin{center}
``V. Summary and interpretation
\end{center}
1. The examined elements \ce{Al}, \ce{Fe}, \ce{Cu}, \ce{Pb} all emit the same wavelength  $\lambda = (\num[output-decimal-marker={.},separate-uncertainty]{23.8 +- 1.0})\, \text{xu}$.

This effect can be interpreted according to Dirac's theory. As L. Meitner and K. Philipp, as well as others (using a cloud chamber), showed, positive electrons occur if hard \ce{Th}C''--$\gamma$ radiation is absorbed, especially in the case of heavy atoms. According to Dirac, the positron is only short--lived ($\sim \num{e-10}\, \text{s}$), because it seeks to recombine with a negative electron, emitting the energy of $2m_0c^2$ during this process. This energy is emitted as two quanta $h\nu = m_0c^2$, that is, the wavelength $\Lambda = \num[output-decimal-marker={.}]{24.2}\,\text{xu}$ will be observed.

2. The intensity of this ``recombination radiation'' per nucles scales with the square of the atomic number. This observation, too, is in accord with the theory. [...]''
\end{quote}

In the months following the publication, \citeauthor{Heiting1934}'s findings were criticized by \citet{BotheHorn1934a}, who doubted that the results of \citet{GrayTarrant1932} and \citet{Heiting1934} were proof of a real secondary radiation of \SI{500}{\kilo\electronvolt}, but they did not rule out the possibility entirely. \citet{BotheHorn1934a} concluded that processes assumed by \citet{MeitnerHupfeld1931}, \citet{GrayTarrant1932}, and \citet{Heiting1934} could only play a subordinate role, even though their experimental results were in good agreement with the aforementioned authors' works. What is more, \citet{BotheHorn1934a} attributed the additional radiative component to bremsstrahlung, or to anomalies of the Compton effect. They also hypothesized that the deceleration and annihilation of a positron occurs in one single radiative process \citep{BotheHorn1934a}. However, \citeauthor{BotheHorn1934a}'s results are not spectrally resolved \citep{BotheHorn1934a}. It appears their measurements do not contradict the results of \citet{Heiting1934}.

Neither in \citet{Amaldi1984} nor in a recent overview article by \citet{Goworek2014} are \citeauthor{Heiting1934}'s articles mentioned. Neither were \citeauthor{Heiting1934}'s articles cited by \citeauthor{Joliot1933} nor \citeauthor{Thibaud1933d}, which might be due to the fact that \citeauthor{Heiting1934} published in German, and also because---in retrospect---the titles of his publications were not ideal. That is, \citeauthor{Heiting1934} never used a term like ``positron annihilation'' or similar as part of the title, even though \citeauthor{Heiting1934} quite clearly concluded that his results were proof of Dirac's theory regarding positron annihilation radiation \citep[][``Rekombinationsstrahlung'']{Heiting1934}. One may argue that because the publications cited here were in English, French, and German, it may be possible that the authors were unaware of each other's findings. This does not seem to be the case, though. They frequently cited each other's articles for various reasons, regardless of the language.

\section{Conclusion}
\label{sec:conclusion}
After inspecting the chronology of submissions and the interpretation of results in these submitted manuscripts, I conclude that \citeauthor{Heiting1933a}'s claim \citep{HeitingHeisenberg1952} seems justified. Indeed, Theodor Heiting should receive credit for the experimental discovery of positron annihilation radiation. \citeauthor{Heiting1933a} submitted his first results on \printdate{07.08.1933}, and these were published shortly after \citep{Heiting1933a}. The interpretation of those results followed in a publication submitted on \printdate{28.09.1933} \citep{Heiting1933b}, where he attributed the secondary radiation to positron annihilation---recombination radiation, as \citeauthor{Heiting1933b} put it at the time. His final publication on the topic was received by the Zeitschrift f\"ur Physik on \printdate{02.11.1933}, which contained a more detailed description of the method and comprehensive experimental results \citep{Heiting1934}. The interpretation of his results remained the same. \citet{Joliot1933}, \citet{Thibaud1933d}, and \citet{Klemperer1934} published their results after \citeauthor{Heiting1933a} published his discovery.  \citeauthor{Heiting1934}'s remark that he was the first to experimentally discover positron annihilation radiation \citep{HeitingHeisenberg1952} seems justified.

\begin{acknowledgements}
I am very grateful to Karin Keller, Universit\"atsarchiv der Universit\"atsbibliothek, and Reinhard Krause--Rehberg, Institut f\"ur Exeprimentalphysik, both of Martin--Luther--Universit\"at Halle--Wittenberg, for their kind help in finding information on Theodor Heiting and sharing these with me, as well as for comments on a draft version. Isabelle Maurin--Joffre kindly provided me with information on the submission and publication procedures in Comptes Rendus. I also thank Eirik Malinen of the Department of Physics, University of Oslo, for helpful discussions and comments on an earlier version of this manuscript.

\end{acknowledgements}

\clearpage
\appendix
\section{Rudimentary biographical notes on Theodor Heiting}
\label{ap:bio}
The following notes are mostly based on archived information at Universit\"atsarchiv der Universit\"atsbibliothek, Martin--Luther--Universit\"at Halle--Wittenberg, Germany. 

Theodor Fritz Heiting was born in Wuppertal--Elberfeld on \printdate{14.07.1908}. His parents were Heinrich and Maria Heiting. He went to the Volksschule in Elberfeld from 1914 until 1918, and went on with a humanistic secondary school education \citep{LMU1928}, at first in Elberfeld until 1919, then eight more years in Halle an der Saale, where he passed his ``A--levels'' in 1927. He went on to study Physics, Mathematics, and Chemistry in Halle with a short intermezzo in Munich. At least in early 1934, he lived at Johannesplatz 14 in Halle.

In Munich, he was a student at Philosophische Fakult\"at, Ludwig--Maximilians--Universit\"at M\"unchen, in winter semester 1928\slash1929 and summer semester 1929. He lived in Augustenstra\ss e 50\slash4 \citep{LMU1928} and Leopoldstra\ss e 60\slash2 \citep{LMU1929}.  At Ludwig--Maximilians--Universit\"at, he took the following courses (teacher in parentheses): introduction to theoretical physics I and II (Graetz), theory of electricity and magnetism (Graetz), thermodynamics and statistical gas theory (Ott), lab course in physics (Kirchner), and differential equations (Perron).

He then continued his studies at Halle--Wittenberg. From September 1931 until July 1933, he was a doctoral candidate with Gerhard Hoffmann. From his early doctoral work, there is an abstract published in Chemisches Zentralblatt \citet{Heiting1932}. On \printdate{06.01.1934}, he applied for his doctoral defence, majoring in physics. His two minors were mathematics and chemistry. Together with his application, he submitted an abstract of his dissertation to the university. I have translated the German abstract, also dated \printdate{06.01.1934}, as follows:
\begin{quote}
``An anomalous absorption is observed as hard $\gamma$ rays pass through heavy atoms. This anomalous absorption cannot be explained by photoelectric nor by Compton scattering processes. Simulatenously, there is a secondary radiation emitted by the irradiated materials which is different from Compton scattering radiation. The investigation of this secondary radiation by C. Y. Chao, L. H. Gray und G. T. P. Tarrant, L. Meitner and H. H. Hupfeld has led to contradictory results.

The purpose of my work was the resolve these contradictions. The secondary radiation of aluminium, iron, copper, and lead was investigated. The radiation of these metals proved to be homogeneous, with the wavelength \SI[output-decimal-marker={.}]{23.8(10)e-11}{\centi\metre} being independent of the material (within the measurement uncertainties), and the intensity of the radiation increased with the square of the atomic number. (Only in lead a second component was found, with a wavelength of \SI[output-decimal-marker={.}]{6.6e-11}{\centi\metre})

This finding was interpreted as follows: If hard $\gamma$ rays interact with matter, a split of the $\gamma$ rays in pairs of positive and negative electrons is possible, and the scattering cross section increases with the square of the atomic number. According to Dirac, the positron is an unstable particle analogous to the photon, it ``recombines'', after it has lost its kinetic energy, with a negative electron, thereby emitting the energy \SI[output-decimal-marker={.}]{1.02e6}{\electronvolt}. From the conservation of momentum, it follows that this energy has two be emitted as two quanta $h\nu = \SI[output-decimal-marker={.}]{0.51e6}{\electronvolt}$, which corresponds to the wavelength \SI[output-decimal-marker={.}]{24.2e-11}{\centi\metre}.''
\end{quote}

On \printdate{28.02.1934}, he passed the doctoral examination at Martin--Luther--Universit\"at Halle--Wittenberg (see footnote \ref{fn:uni} on page \pageref{fn:uni}), and was awarded a doctoral degree in science (Dr. rer. nat.) on \printdate{08.03.1934} for his dissertation entitled ``Untersuchungen \"uber die durch harte $\gamma$--Strahlung hervorgerufene Sekund\"arstrahlung''\footnote{Translated as ``Investigations of the secondary radiation due to hard $\gamma$ rays''}.

\clearpage
\bibliography{positron.bib}

\begin{thebibliography}{51}%
\makeatletter
\providecommand \@ifxundefined [1]{%
 \@ifx{#1\undefined}
}%
\providecommand \@ifnum [1]{%
 \ifnum #1\expandafter \@firstoftwo
 \else \expandafter \@secondoftwo
 \fi
}%
\providecommand \@ifx [1]{%
 \ifx #1\expandafter \@firstoftwo
 \else \expandafter \@secondoftwo
 \fi
}%
\providecommand \natexlab [1]{#1}%
\providecommand \enquote  [1]{``#1''}%
\providecommand \bibnamefont  [1]{#1}%
\providecommand \bibfnamefont [1]{#1}%
\providecommand \citenamefont [1]{#1}%
\providecommand \href@noop [0]{\@secondoftwo}%
\providecommand \href [0]{\begingroup \@sanitize@url \@href}%
\providecommand \@href[1]{\@@startlink{#1}\@@href}%
\providecommand \@@href[1]{\endgroup#1\@@endlink}%
\providecommand \@sanitize@url [0]{\catcode `\\12\catcode `\$12\catcode
  `\&12\catcode `\#12\catcode `\^12\catcode `\_12\catcode `\%12\relax}%
\providecommand \@@startlink[1]{}%
\providecommand \@@endlink[0]{}%
\providecommand \url  [0]{\begingroup\@sanitize@url \@url }%
\providecommand \@url [1]{\endgroup\@href {#1}{\urlprefix }}%
\providecommand \urlprefix  [0]{URL }%
\providecommand \Eprint [0]{\href }%
\providecommand \doibase [0]{http://dx.doi.org/}%
\providecommand \selectlanguage [0]{\@gobble}%
\providecommand \bibinfo  [0]{\@secondoftwo}%
\providecommand \bibfield  [0]{\@secondoftwo}%
\providecommand \translation [1]{[#1]}%
\providecommand \BibitemOpen [0]{}%
\providecommand \bibitemStop [0]{}%
\providecommand \bibitemNoStop [0]{.\EOS\space}%
\providecommand \EOS [0]{\spacefactor3000\relax}%
\providecommand \BibitemShut  [1]{\csname bibitem#1\endcsname}%
\let\auto@bib@innerbib\@empty
\bibitem [{\citenamefont {{Al--Ramadhan}}\ and\ \citenamefont
  {Gidley}(1994)}]{Al-RamadhanGidley1994}%
  \BibitemOpen
  \bibfield  {author} {\bibinfo {author} {\bibnamefont {{Al--Ramadhan}},
  \bibfnamefont {A~H}}, \ and\ \bibinfo {author} {\bibfnamefont {D.~W.}\
  \bibnamefont {Gidley}}} (\bibinfo {year} {1994}),\ \bibfield  {title}
  {\enquote {\bibinfo {title} {New precision measurement of the decay rate of
  singlet positronium},}\ }\href {\doibase 10.1103/PhysRevLett.72.1632}
  {\bibfield  {journal} {\bibinfo  {journal} {Phys. Rev. Lett.}\ }\textbf
  {\bibinfo {volume} {72}}~(\bibinfo {number} {11}),\ \bibinfo {pages}
  {1632--1635}}\BibitemShut {NoStop}%
\bibitem [{\citenamefont {Amaldi}(1984)}]{Amaldi1984}%
  \BibitemOpen
  \bibfield  {author} {\bibinfo {author} {\bibnamefont {Amaldi}, \bibfnamefont
  {E}}} (\bibinfo {year} {1984}),\ \bibfield  {title} {\enquote {\bibinfo
  {title} {From the discovery of the neutron to the discovery of nuclear
  fission},}\ }\href {\doibase 10.1016/0370-1573(84)90214-X} {\bibfield
  {journal} {\bibinfo  {journal} {Phys. Rep.}\ }\textbf {\bibinfo {volume}
  {111}}~(\bibinfo {number} {1}),\ \bibinfo {pages} {1--331}}\BibitemShut
  {NoStop}%
\bibitem [{\citenamefont {Anderson}(1932)}]{Anderson1932}%
  \BibitemOpen
  \bibfield  {author} {\bibinfo {author} {\bibnamefont {Anderson},
  \bibfnamefont {C~D}}} (\bibinfo {year} {1932}),\ \bibfield  {title} {\enquote
  {\bibinfo {title} {The apparent existence of easily deflectable positives},}\
  }\href {\doibase 10.1126/science.76.1967.238} {\bibfield  {journal} {\bibinfo
   {journal} {Science}\ }\textbf {\bibinfo {volume} {76}}~(\bibinfo {number}
  {1967}),\ \bibinfo {pages} {238--239}}\BibitemShut {NoStop}%
\bibitem [{\citenamefont {Anderson}(1933{\natexlab{a}})}]{Anderson1933b}%
  \BibitemOpen
  \bibfield  {author} {\bibinfo {author} {\bibnamefont {Anderson},
  \bibfnamefont {C~D}}} (\bibinfo {year} {1933}{\natexlab{a}}),\ \bibfield
  {title} {\enquote {\bibinfo {title} {Cosmic--{R}ay {P}ositive and {N}egative
  {E}lectrons},}\ }\href {\doibase 10.1103/PhysRev.44.406} {\bibfield
  {journal} {\bibinfo  {journal} {Phys. Rev.}\ }\textbf {\bibinfo {volume}
  {44}}~(\bibinfo {number} {5}),\ \bibinfo {pages} {406--416}}\BibitemShut
  {NoStop}%
\bibitem [{\citenamefont {Anderson}(1933{\natexlab{b}})}]{Anderson1933a}%
  \BibitemOpen
  \bibfield  {author} {\bibinfo {author} {\bibnamefont {Anderson},
  \bibfnamefont {C~D}}} (\bibinfo {year} {1933}{\natexlab{b}}),\ \bibfield
  {title} {\enquote {\bibinfo {title} {The {P}ositive {E}lectron},}\ }\href
  {\doibase 10.1103/PhysRev.43.491} {\bibfield  {journal} {\bibinfo  {journal}
  {Phys. Rev.}\ }\textbf {\bibinfo {volume} {43}}~(\bibinfo {number} {6}),\
  \bibinfo {pages} {491--494}}\BibitemShut {NoStop}%
\bibitem [{\citenamefont {Anderson}\ and\ \citenamefont
  {Weiner}(1966)}]{AndersonWeiner1966}%
  \BibitemOpen
  \bibfield  {author} {\bibinfo {author} {\bibnamefont {Anderson},
  \bibfnamefont {C~D}}, \ and\ \bibinfo {author} {\bibfnamefont
  {C.}~\bibnamefont {Weiner}}} (\bibinfo {year} {1966}),\ \href
  {https://www.aip.org/history-programs/niels-bohr-library/oral-histories/4487}
  {\enquote {\bibinfo {title} {Interview of {C}arl {A}nderson by {C}harles
  {W}einer on 1966 {J}une 30},}\ }\bibinfo {howpublished} {Niels Bohr Library
  \& Archives, American Institute of Physics, College Park, MD USA}\BibitemShut
  {NoStop}%
\bibitem [{\citenamefont {Bearden}(1965)}]{Bearden1965}%
  \BibitemOpen
  \bibfield  {author} {\bibinfo {author} {\bibnamefont {Bearden}, \bibfnamefont
  {J~A}}} (\bibinfo {year} {1965}),\ \bibfield  {title} {\enquote {\bibinfo
  {title} {X--{R}ay {W}avelength {C}onversion {F}actor
  $\lambda\left(\lambda_g/\lambda_s\right)^*$},}\ }\href {\doibase
  10.1103/PhysRev.137.B181} {\bibfield  {journal} {\bibinfo  {journal} {Phys.
  Rev.}\ }\textbf {\bibinfo {volume} {137}}~(\bibinfo {number} {1B}),\ \bibinfo
  {pages} {B181--B187}}\BibitemShut {NoStop}%
\bibitem [{\citenamefont {Blackett}\ and\ \citenamefont
  {Occhialini}(1933)}]{BlackettOcchialini1933}%
  \BibitemOpen
  \bibfield  {author} {\bibinfo {author} {\bibnamefont {Blackett},
  \bibfnamefont {P~M~S}}, \ and\ \bibinfo {author} {\bibfnamefont {G.~P.~S.}\
  \bibnamefont {Occhialini}}} (\bibinfo {year} {1933}),\ \bibfield  {title}
  {\enquote {\bibinfo {title} {Some {P}hotographs of the {T}racks of
  {P}enetrating {R}adiation},}\ }\href {\doibase 10.1098/rspa.1933.0048}
  {\bibfield  {journal} {\bibinfo  {journal} {Proc. R. Soc. Lond. A}\ }\textbf
  {\bibinfo {volume} {139}}~(\bibinfo {number} {839}),\ \bibinfo {pages}
  {699--726}}\BibitemShut {NoStop}%
\bibitem [{\citenamefont {Bothe}\ and\ \citenamefont
  {Horn}(1934{\natexlab{a}})}]{BotheHorn1934a}%
  \BibitemOpen
  \bibfield  {author} {\bibinfo {author} {\bibnamefont {Bothe}, \bibfnamefont
  {W}}, \ and\ \bibinfo {author} {\bibfnamefont {W.}~\bibnamefont {Horn}}}
  (\bibinfo {year} {1934}{\natexlab{a}}),\ \bibfield  {title} {\enquote
  {\bibinfo {title} {Die {S}ekund\"arstrahlung harter $\gamma$--{S}trahlen},}\
  }\href {\doibase 10.1007/BF01495384} {\bibfield  {journal} {\bibinfo
  {journal} {Naturwissenschaften}\ }\textbf {\bibinfo {volume} {22}}~(\bibinfo
  {number} {7}),\ \bibinfo {pages} {106--107}},\ \bibinfo {note} {in
  German}\BibitemShut {NoStop}%
\bibitem [{\citenamefont {Bothe}\ and\ \citenamefont
  {Horn}(1934{\natexlab{b}})}]{BotheHorn1934b}%
  \BibitemOpen
  \bibfield  {author} {\bibinfo {author} {\bibnamefont {Bothe}, \bibfnamefont
  {W}}, \ and\ \bibinfo {author} {\bibfnamefont {W.}~\bibnamefont {Horn}}}
  (\bibinfo {year} {1934}{\natexlab{b}}),\ \bibfield  {title} {\enquote
  {\bibinfo {title} {Die {S}ekund\"arstrahlung harter $\gamma$--{S}trahlen},}\
  }\href {\doibase 10.1007/BF01333117} {\bibfield  {journal} {\bibinfo
  {journal} {Z. Phys.}\ }\textbf {\bibinfo {volume} {88}}~(\bibinfo {number}
  {9--10}),\ \bibinfo {pages} {683--698}},\ \bibinfo {note} {in
  German}\BibitemShut {NoStop}%
\bibitem [{\citenamefont {Br\"uche}\ \emph {et~al.}(1966)\citenamefont
  {Br\"uche}, \citenamefont {Bethge},\ and\ \citenamefont
  {Schmidt}}]{Bruecheetal1966}%
  \BibitemOpen
  \bibfield  {author} {\bibinfo {author} {\bibnamefont {Br\"uche},
  \bibfnamefont {E}}, \bibinfo {author} {\bibfnamefont {H.}~\bibnamefont
  {Bethge}}, \ and\ \bibinfo {author} {\bibfnamefont {G.}~\bibnamefont
  {Schmidt}}} (\bibinfo {year} {1966}),\ \bibfield  {title} {\enquote {\bibinfo
  {title} {Maximilian {P}fl\"ucke/{W}ilhelm {M}esserschmidt 60 {J}ahre},}\
  }\href {\doibase 10.1002/phbl.19660220306} {\bibfield  {journal} {\bibinfo
  {journal} {Phys. Bl.}\ }\textbf {\bibinfo {volume} {22}}~(\bibinfo {number}
  {3}),\ \bibinfo {pages} {133--134}}\BibitemShut {NoStop}%
\bibitem [{\citenamefont {Chadwick}\ \emph {et~al.}(1933)\citenamefont
  {Chadwick}, \citenamefont {Blackett},\ and\ \citenamefont
  {Occhialini}}]{Chadwicketal1934}%
  \BibitemOpen
  \bibfield  {author} {\bibinfo {author} {\bibnamefont {Chadwick},
  \bibfnamefont {J}}, \bibinfo {author} {\bibfnamefont {P.~M.~S.}\ \bibnamefont
  {Blackett}}, \ and\ \bibinfo {author} {\bibfnamefont {G.~P.~S.}\ \bibnamefont
  {Occhialini}}} (\bibinfo {year} {1933}),\ \bibfield  {title} {\enquote
  {\bibinfo {title} {Some {E}xperiments on the {P}roduction of {P}ositive
  {E}lectrons},}\ }\href {\doibase 10.1098/rspa.1934.0045} {\bibfield
  {journal} {\bibinfo  {journal} {Proc. R. Soc. Lond. A}\ }\textbf {\bibinfo
  {volume} {144}}~(\bibinfo {number} {851}),\ \bibinfo {pages}
  {235--249}}\BibitemShut {NoStop}%
\bibitem [{\citenamefont {Chao}(1930)}]{Chao1930}%
  \BibitemOpen
  \bibfield  {author} {\bibinfo {author} {\bibnamefont {Chao}, \bibfnamefont
  {C~Y}}} (\bibinfo {year} {1930}),\ \bibfield  {title} {\enquote {\bibinfo
  {title} {Scattering of hard $\gamma$--rays},}\ }\href {\doibase
  10.1103/PhysRev.36.1519} {\bibfield  {journal} {\bibinfo  {journal} {Phys.
  Rev.}\ }\textbf {\bibinfo {volume} {36}}~(\bibinfo {number} {10}),\ \bibinfo
  {pages} {1519--1522}}\BibitemShut {NoStop}%
\bibitem [{\citenamefont {Crane}\ and\ \citenamefont
  {Lauritsen}(1934)}]{CraneLauritsen1934}%
  \BibitemOpen
  \bibfield  {author} {\bibinfo {author} {\bibnamefont {Crane}, \bibfnamefont
  {H~R}}, \ and\ \bibinfo {author} {\bibfnamefont {C.~C.}\ \bibnamefont
  {Lauritsen}}} (\bibinfo {year} {1934}),\ \bibfield  {title} {\enquote
  {\bibinfo {title} {Radioactivity from {C}arbon and {B}oron {O}xide
  {B}ombarded with {D}eutons and the {C}onversion of {P}ositrons into
  {R}adiation},}\ }\href {\doibase 10.1103/PhysRev.45.430.2} {\bibfield
  {journal} {\bibinfo  {journal} {Phys. Rev.}\ }\textbf {\bibinfo {volume}
  {45}}~(\bibinfo {number} {6}),\ \bibinfo {pages} {430--432}}\BibitemShut
  {NoStop}%
\bibitem [{\citenamefont {Deutsch}(1951)}]{Deutsch1951}%
  \BibitemOpen
  \bibfield  {author} {\bibinfo {author} {\bibnamefont {Deutsch}, \bibfnamefont
  {M}}} (\bibinfo {year} {1951}),\ \bibfield  {title} {\enquote {\bibinfo
  {title} {Evidence for the {F}ormation of {P}ositronium in {G}ases},}\ }\href
  {\doibase 10.1103/PhysRev.82.455} {\bibfield  {journal} {\bibinfo  {journal}
  {Phys. Rev.}\ }\textbf {\bibinfo {volume} {82}}~(\bibinfo {number} {3}),\
  \bibinfo {pages} {455--456}}\BibitemShut {NoStop}%
\bibitem [{\citenamefont {Dirac}(1928)}]{Dirac1928}%
  \BibitemOpen
  \bibfield  {author} {\bibinfo {author} {\bibnamefont {Dirac}, \bibfnamefont
  {P~A~M}}} (\bibinfo {year} {1928}),\ \bibfield  {title} {\enquote {\bibinfo
  {title} {On the {Q}uantum {T}heory of the {E}lectron},}\ }\href {\doibase
  10.1098/rspa.1928.0023} {\bibfield  {journal} {\bibinfo  {journal} {Proc. R.
  Soc. Lond. A}\ }\textbf {\bibinfo {volume} {117}}~(\bibinfo {number} {778}),\
  \bibinfo {pages} {610--624}}\BibitemShut {NoStop}%
\bibitem [{\citenamefont {Dirac}(1930)}]{Dirac1930}%
  \BibitemOpen
  \bibfield  {author} {\bibinfo {author} {\bibnamefont {Dirac}, \bibfnamefont
  {P~A~M}}} (\bibinfo {year} {1930}),\ \bibfield  {title} {\enquote {\bibinfo
  {title} {On the {A}nnihilation of {E}lectrons and {P}rotons},}\ }\href
  {\doibase 10.1017/S0305004100016091} {\bibfield  {journal} {\bibinfo
  {journal} {Math. Proc. Cambridge}\ }\textbf {\bibinfo {volume}
  {26}}~(\bibinfo {number} {3}),\ \bibinfo {pages} {361--375}}\BibitemShut
  {NoStop}%
\bibitem [{\citenamefont {Dirac}(1931)}]{Dirac1931}%
  \BibitemOpen
  \bibfield  {author} {\bibinfo {author} {\bibnamefont {Dirac}, \bibfnamefont
  {P~A~M}}} (\bibinfo {year} {1931}),\ \bibfield  {title} {\enquote {\bibinfo
  {title} {Quantised {S}ingularities in the {E}lectromagnetic {F}ield},}\
  }\href {\doibase 10.1098/rspa.1931.0130} {\bibfield  {journal} {\bibinfo
  {journal} {Proc. R. Soc. Lond. A}\ }\textbf {\bibinfo {volume}
  {133}}~(\bibinfo {number} {821}),\ \bibinfo {pages} {60--72}}\BibitemShut
  {NoStop}%
\bibitem [{\citenamefont {Dirac}(1933)}]{Dirac1933}%
  \BibitemOpen
  \bibfield  {author} {\bibinfo {author} {\bibnamefont {Dirac}, \bibfnamefont
  {P~A~M}}} (\bibinfo {year} {1933}),\ \href
  {http://www.nobelprize.org/nobel_prizes/physics/laureates/1933/dirac-lecture.html}
  {\enquote {\bibinfo {title} {Theory of {E}lectrons and {P}ositrons},}\
  }\bibinfo {howpublished} {Nobel lecture},\ \bibinfo {note} {accessed: 12
  October 2014}\BibitemShut {NoStop}%
\bibitem [{\citenamefont {Fermi}\ and\ \citenamefont
  {Uhlenbeck}(1933)}]{FermiUhlenbeck1934}%
  \BibitemOpen
  \bibfield  {author} {\bibinfo {author} {\bibnamefont {Fermi}, \bibfnamefont
  {E}}, \ and\ \bibinfo {author} {\bibfnamefont {G.~E.}\ \bibnamefont
  {Uhlenbeck}}} (\bibinfo {year} {1933}),\ \bibfield  {title} {\enquote
  {\bibinfo {title} {On the {R}ecombination of {E}lectrons and {P}ositrons},}\
  }\href {\doibase 10.1103/PhysRev.45.430.2} {\bibfield  {journal} {\bibinfo
  {journal} {Phys. Rev.}\ }\textbf {\bibinfo {volume} {44}}~(\bibinfo {number}
  {6}),\ \bibinfo {pages} {510--511}}\BibitemShut {NoStop}%
\bibitem [{\citenamefont {Fleischmann}\ and\ \citenamefont
  {Bothe}(1934)}]{FleischmannBothe1934}%
  \BibitemOpen
  \bibfield  {author} {\bibinfo {author} {\bibnamefont {Fleischmann},
  \bibfnamefont {R}}, \ and\ \bibinfo {author} {\bibfnamefont {W.}~\bibnamefont
  {Bothe}}} (\bibinfo {year} {1934}),\ \enquote {\bibinfo {title} {K\"unstliche
  {K}ern--$\gamma$--{S}trahlen, {N}eutronen, {P}ositronen},}\ in\ \href
  {\doibase 10.1007/BFb0112003} {\emph {\bibinfo {booktitle} {Ergebnisse der
  exakten Naturwissenschaften}}},\ Vol.~\bibinfo {volume} {13}\ (\bibinfo
  {publisher} {Springer},\ \bibinfo {address} {Berlin})\ pp.\ \bibinfo {pages}
  {1--56},\ \bibinfo {note} {in German}\BibitemShut {NoStop}%
\bibitem [{\citenamefont {Gentner}(1934{\natexlab{a}})}]{Gentner1934a}%
  \BibitemOpen
  \bibfield  {author} {\bibinfo {author} {\bibnamefont {Gentner}, \bibfnamefont
  {W}}} (\bibinfo {year} {1934}{\natexlab{a}}),\ \bibfield  {title} {\enquote
  {\bibinfo {title} {Sur l'absorption des rayon $\gamma$ p\'{e}n\'{e}trants},}\
  }\href {\doibase 10.1051/jphysrad:019340050204900} {\bibfield  {journal}
  {\bibinfo  {journal} {J. Phys. Radium}\ }\textbf {\bibinfo {volume}
  {5}}~(\bibinfo {number} {2}),\ \bibinfo {pages} {49--53}},\ \bibinfo {note}
  {in French}\BibitemShut {NoStop}%
\bibitem [{\citenamefont {Gentner}(1934{\natexlab{b}})}]{Gentner1934b}%
  \BibitemOpen
  \bibfield  {author} {\bibinfo {author} {\bibnamefont {Gentner}, \bibfnamefont
  {W}}} (\bibinfo {year} {1934}{\natexlab{b}}),\ \bibfield  {title} {\enquote
  {\bibinfo {title} {Zur {W}ellenl\"ange und {I}ntensit\"at der
  {S}ekund\"arstrahlung harter $\gamma$--{S}trahlen},}\ }\href {\doibase
  10.1007/BF01495566} {\bibfield  {journal} {\bibinfo  {journal}
  {Naturwissenschaften}\ }\textbf {\bibinfo {volume} {22}}~(\bibinfo {number}
  {25}),\ \bibinfo {pages} {435}},\ \bibinfo {note} {in German}\BibitemShut
  {NoStop}%
\bibitem [{\citenamefont {Goworek}(2014)}]{Goworek2014}%
  \BibitemOpen
  \bibfield  {author} {\bibinfo {author} {\bibnamefont {Goworek}, \bibfnamefont
  {T}}} (\bibinfo {year} {2014}),\ \bibfield  {title} {\enquote {\bibinfo
  {title} {80 {Y}ears of {P}ositron {A}nnihilation {R}adiation},}\ }\href
  {\doibase 10.12693/APhysPolA.125.685} {\bibfield  {journal} {\bibinfo
  {journal} {Acta Phys. Pol. A}\ }\textbf {\bibinfo {volume} {125}}~(\bibinfo
  {number} {3}),\ \bibinfo {pages} {685--687}}\BibitemShut {NoStop}%
\bibitem [{\citenamefont {Gray}\ and\ \citenamefont
  {Tarrant}(1932)}]{GrayTarrant1932}%
  \BibitemOpen
  \bibfield  {author} {\bibinfo {author} {\bibnamefont {Gray}, \bibfnamefont
  {L~H}}, \ and\ \bibinfo {author} {\bibfnamefont {G.~T.~P.}\ \bibnamefont
  {Tarrant}}} (\bibinfo {year} {1932}),\ \bibfield  {title} {\enquote {\bibinfo
  {title} {The {N}ature of the {I}nteraction between {G}amma--{R}adiation and
  the {A}tomic {N}ucleus},}\ }\href {\doibase 10.1098/rspa.1932.0111}
  {\bibfield  {journal} {\bibinfo  {journal} {Proc. R. Soc. Lond. A}\ }\textbf
  {\bibinfo {volume} {136}}~(\bibinfo {number} {830}),\ \bibinfo {pages}
  {662--691}}\BibitemShut {NoStop}%
\bibitem [{\citenamefont {Gray}\ and\ \citenamefont
  {Tarrant}(1934{\natexlab{a}})}]{GrayTarrant1934a}%
  \BibitemOpen
  \bibfield  {author} {\bibinfo {author} {\bibnamefont {Gray}, \bibfnamefont
  {L~H}}, \ and\ \bibinfo {author} {\bibfnamefont {G.~T.~P.}\ \bibnamefont
  {Tarrant}}} (\bibinfo {year} {1934}{\natexlab{a}}),\ \bibfield  {title}
  {\enquote {\bibinfo {title} {Phenomena {A}ssociated with the {A}nomalous
  {A}bsorption of {H}igh {E}nergy {G}amma {R}adiation. -- {II}},}\ }\href
  {\doibase 10.1098/rspa.1934.0028} {\bibfield  {journal} {\bibinfo  {journal}
  {Proc. R. Soc. Lond. A}\ }\textbf {\bibinfo {volume} {143}}~(\bibinfo
  {number} {850}),\ \bibinfo {pages} {681--706}}\BibitemShut {NoStop}%
\bibitem [{\citenamefont {Gray}\ and\ \citenamefont
  {Tarrant}(1934{\natexlab{b}})}]{GrayTarrant1934b}%
  \BibitemOpen
  \bibfield  {author} {\bibinfo {author} {\bibnamefont {Gray}, \bibfnamefont
  {L~H}}, \ and\ \bibinfo {author} {\bibfnamefont {G.~T.~P.}\ \bibnamefont
  {Tarrant}}} (\bibinfo {year} {1934}{\natexlab{b}}),\ \bibfield  {title}
  {\enquote {\bibinfo {title} {Phenomena {A}ssociated with the {A}nomalous
  {A}bsorption of {H}igh {E}nergy {G}amma {R}adiation. -- {III}},}\ }\href
  {\doibase 10.1098/rspa.1934.0029} {\bibfield  {journal} {\bibinfo  {journal}
  {Proc. R. Soc. Lond. A}\ }\textbf {\bibinfo {volume} {143}}~(\bibinfo
  {number} {850}),\ \bibinfo {pages} {706--724}}\BibitemShut {NoStop}%
\bibitem [{\citenamefont {Guerra}\ \emph {et~al.}(2012)\citenamefont {Guerra},
  \citenamefont {Leone},\ and\ \citenamefont {Robotti}}]{Guerraetal2012}%
  \BibitemOpen
  \bibfield  {author} {\bibinfo {author} {\bibnamefont {Guerra}, \bibfnamefont
  {F}}, \bibinfo {author} {\bibfnamefont {M.}~\bibnamefont {Leone}}, \ and\
  \bibinfo {author} {\bibfnamefont {N.}~\bibnamefont {Robotti}}} (\bibinfo
  {year} {2012}),\ \bibfield  {title} {\enquote {\bibinfo {title} {The
  {D}iscovery of {A}rtifical {R}adioactivity},}\ }\href {\doibase
  10.1007/s00016-011-0064-7} {\bibfield  {journal} {\bibinfo  {journal} {Phys.
  Perspect.}\ }\textbf {\bibinfo {volume} {14}},\ \bibinfo {pages}
  {33--58}}\BibitemShut {NoStop}%
\bibitem [{\citenamefont {Hanson}(1961)}]{Hanson1961}%
  \BibitemOpen
  \bibfield  {author} {\bibinfo {author} {\bibnamefont {Hanson}, \bibfnamefont
  {N~R}}} (\bibinfo {year} {1961}),\ \bibfield  {title} {\enquote {\bibinfo
  {title} {Discovering the positron ({I})$^*$},}\ }\href {\doibase
  10.1093/bjps/XII.47.194} {\bibfield  {journal} {\bibinfo  {journal} {Br. J.
  Phil. Sci.}\ }\textbf {\bibinfo {volume} {XII}}~(\bibinfo {number} {47}),\
  \bibinfo {pages} {194--214}}\BibitemShut {NoStop}%
\bibitem [{\citenamefont {Heiting}(1932)}]{Heiting1932}%
  \BibitemOpen
  \bibfield  {author} {\bibinfo {author} {\bibnamefont {Heiting}, \bibfnamefont
  {T}}} (\bibinfo {year} {1932}),\ \bibfield  {title} {\enquote {\bibinfo
  {title} {Vergleich einer {A}rbeit von {G}. {H}offmann \"uber den
  {C}omptoneffekt bei $\gamma$--{S}trahlen mit der neueren {T}heorie},}\
  }\href@noop {} {\bibfield  {journal} {\bibinfo  {journal} {Chemisches
  Zentralblatt}\ }\textbf {\bibinfo {volume} {I}}~(\bibinfo {number} {23}),\
  \bibinfo {pages} {3032}},\ \bibinfo {note} {in German}\BibitemShut {NoStop}%
\bibitem [{\citenamefont {Heiting}(1933{\natexlab{a}})}]{Heiting1933a}%
  \BibitemOpen
  \bibfield  {author} {\bibinfo {author} {\bibnamefont {Heiting}, \bibfnamefont
  {T}}} (\bibinfo {year} {1933}{\natexlab{a}}),\ \bibfield  {title} {\enquote
  {\bibinfo {title} {Kernanregung durch harte $\gamma$--{S}trahlen},}\ }\href
  {\doibase 10.1007/BF01504047} {\bibfield  {journal} {\bibinfo  {journal}
  {Naturwissenschaften}\ }\textbf {\bibinfo {volume} {21}}~(\bibinfo {number}
  {37}),\ \bibinfo {pages} {674}},\ \bibinfo {note} {in German}\BibitemShut
  {NoStop}%
\bibitem [{\citenamefont {Heiting}(1933{\natexlab{b}})}]{Heiting1933b}%
  \BibitemOpen
  \bibfield  {author} {\bibinfo {author} {\bibnamefont {Heiting}, \bibfnamefont
  {T}}} (\bibinfo {year} {1933}{\natexlab{b}}),\ \bibfield  {title} {\enquote
  {\bibinfo {title} {Zur {K}ern--$\gamma$--{A}bsorption},}\ }\href {\doibase
  10.1007/BF01505054} {\bibfield  {journal} {\bibinfo  {journal}
  {Naturwissenschaften}\ }\textbf {\bibinfo {volume} {21}}~(\bibinfo {number}
  {45}),\ \bibinfo {pages} {800}},\ \bibinfo {note} {in German}\BibitemShut
  {NoStop}%
\bibitem [{\citenamefont {Heiting}(1934)}]{Heiting1934}%
  \BibitemOpen
  \bibfield  {author} {\bibinfo {author} {\bibnamefont {Heiting}, \bibfnamefont
  {T}}} (\bibinfo {year} {1934}),\ \bibfield  {title} {\enquote {\bibinfo
  {title} {Untersuchung \"uber die durch harte $\gamma$--{S}trahlung
  hervorgerufene {S}ekund\"arstrahlung},}\ }\href {\doibase 10.1007/BF01338454}
  {\bibfield  {journal} {\bibinfo  {journal} {Z. Phys.}\ }\textbf {\bibinfo
  {volume} {87}}~(\bibinfo {number} {1--2}),\ \bibinfo {pages} {127--138}},\
  \bibinfo {note} {in German}\BibitemShut {NoStop}%
\bibitem [{\citenamefont {Heiting}\ and\ \citenamefont
  {Heisenberg}(1952)}]{HeitingHeisenberg1952}%
  \BibitemOpen
  \bibfield  {author} {\bibinfo {author} {\bibnamefont {Heiting}, \bibfnamefont
  {T}}, \ and\ \bibinfo {author} {\bibfnamefont {W.}~\bibnamefont
  {Heisenberg}}} (\bibinfo {year} {1952}),\ \bibfield  {title} {\enquote
  {\bibinfo {title} {Zur {Z}erstrahlung von {E}lektron und {P}ositron},}\
  }\href {\doibase 10.1002/phbl.19520080410} {\bibfield  {journal} {\bibinfo
  {journal} {Phys. Bl.}\ }\textbf {\bibinfo {volume} {8}}~(\bibinfo {number}
  {4}),\ \bibinfo {pages} {191}},\ \bibinfo {note} {in German}\BibitemShut
  {NoStop}%
\bibitem [{\citenamefont {Hill}\ and\ \citenamefont
  {Landshoff}(1938)}]{HillLandshoff1938}%
  \BibitemOpen
  \bibfield  {author} {\bibinfo {author} {\bibnamefont {Hill}, \bibfnamefont
  {E~L}}, \ and\ \bibinfo {author} {\bibfnamefont {R.}~\bibnamefont
  {Landshoff}}} (\bibinfo {year} {1938}),\ \bibfield  {title} {\enquote
  {\bibinfo {title} {The {D}irac {E}lectron {T}heory},}\ }\href {\doibase
  10.1103/RevModPhys.10.87} {\bibfield  {journal} {\bibinfo  {journal} {Rev.
  Mod. Phys.}\ }\textbf {\bibinfo {volume} {10}}~(\bibinfo {number} {2}),\
  \bibinfo {pages} {87--132}}\BibitemShut {NoStop}%
\bibitem [{\citenamefont {Ising}(1964)}]{Ising1964}%
  \BibitemOpen
  \bibfield  {author} {\bibinfo {author} {\bibnamefont {Ising}, \bibfnamefont
  {G}}} (\bibinfo {year} {1964}),\ \enquote {\bibinfo {title} {Physics 1948.
  {P}resentation {S}peech by {P}rofessor {G}. {I}sing, member of the {N}obel
  {C}ommittee for {P}hysics},}\ in\ \href {\doibase
  10.1016/B978-1-4831-9746-3.50011-9} {\emph {\bibinfo {booktitle} {Physics
  1942--1962}}}\ (\bibinfo  {publisher} {Elsevier},\ \bibinfo {address}
  {Amsterdam--London--New York})\ pp.\ \bibinfo {pages} {93--96}\BibitemShut
  {NoStop}%
\bibitem [{\citenamefont {Joliot}(1933)}]{Joliot1933}%
  \BibitemOpen
  \bibfield  {author} {\bibinfo {author} {\bibnamefont {Joliot}, \bibfnamefont
  {F}}} (\bibinfo {year} {1933}),\ \bibfield  {title} {\enquote {\bibinfo
  {title} {Preuve exp\'{e}rimentale de l'annihilation des \'{e}lectrons
  positifs},}\ }\href@noop {} {\bibfield  {journal} {\bibinfo  {journal} {C. R.
  Hebd. Acad. Sci.}\ }\textbf {\bibinfo {volume} {197}}~(\bibinfo {number}
  {25}),\ \bibinfo {pages} {1622--1625}},\ \bibinfo {note} {in
  French}\BibitemShut {NoStop}%
\bibitem [{\citenamefont {Joliot}(1934{\natexlab{a}})}]{Joliot1934b}%
  \BibitemOpen
  \bibfield  {author} {\bibinfo {author} {\bibnamefont {Joliot}, \bibfnamefont
  {F}}} (\bibinfo {year} {1934}{\natexlab{a}}),\ \bibfield  {title} {\enquote
  {\bibinfo {title} {Preuve exp\'{e}rimentale de l'annihilation des
  \'{e}lectrons positifs},}\ }\href {\doibase
  10.1051/jphysrad:0193400507029900} {\bibfield  {journal} {\bibinfo  {journal}
  {J. Phys. Radium}\ }\textbf {\bibinfo {volume} {5}}~(\bibinfo {number} {7}),\
  \bibinfo {pages} {299--303}},\ \bibinfo {note} {in French}\BibitemShut
  {NoStop}%
\bibitem [{\citenamefont {Joliot}(1934{\natexlab{b}})}]{Joliot1934a}%
  \BibitemOpen
  \bibfield  {author} {\bibinfo {author} {\bibnamefont {Joliot}, \bibfnamefont
  {F}}} (\bibinfo {year} {1934}{\natexlab{b}}),\ \bibfield  {title} {\enquote
  {\bibinfo {title} {Sur la d\'{e}mat\'{e}ralisation de paires
  d'\'{e}lectrons},}\ }\href@noop {} {\bibfield  {journal} {\bibinfo  {journal}
  {C. R. Hebd. Acad. Sci.}\ }\textbf {\bibinfo {volume} {198}}~(\bibinfo
  {number} {26}),\ \bibinfo {pages} {81--83}},\ \bibinfo {note} {in
  French}\BibitemShut {NoStop}%
\bibitem [{\citenamefont {Klemperer}(1934)}]{Klemperer1934}%
  \BibitemOpen
  \bibfield  {author} {\bibinfo {author} {\bibnamefont {Klemperer},
  \bibfnamefont {O}}} (\bibinfo {year} {1934}),\ \bibfield  {title} {\enquote
  {\bibinfo {title} {On the annihilation radiation of the positron},}\ }\href
  {\doibase 10.1017/S0305004100012536} {\bibfield  {journal} {\bibinfo
  {journal} {Math. Proc. Cambridge}\ }\textbf {\bibinfo {volume}
  {30}}~(\bibinfo {number} {3}),\ \bibinfo {pages} {347--354}}\BibitemShut
  {NoStop}%
\bibitem [{\citenamefont {Leone}\ and\ \citenamefont
  {Robotti}(2012)}]{LeoneRobotti2012}%
  \BibitemOpen
  \bibfield  {author} {\bibinfo {author} {\bibnamefont {Leone}, \bibfnamefont
  {M}}, \ and\ \bibinfo {author} {\bibfnamefont {N.}~\bibnamefont {Robotti}}}
  (\bibinfo {year} {2012}),\ \bibfield  {title} {\enquote {\bibinfo {title} {An
  uninvited guest: {T}he positron in early 1930s physics},}\ }\href {\doibase
  10.1119/1.3695374} {\bibfield  {journal} {\bibinfo  {journal} {Am. J. Phys.}\
  }\textbf {\bibinfo {volume} {80}}~(\bibinfo {number} {6}),\ \bibinfo {pages}
  {534--541}}\BibitemShut {NoStop}%
\bibitem [{\citenamefont {{LMU M\"unchen}}(1928)}]{LMU1928}%
  \BibitemOpen
  \bibfield  {author} {\bibinfo {author} {\bibnamefont {{LMU M\"unchen}},}}
  (\bibinfo {year} {1928}),\ \href
  {http://nbn-resolving.de/urn/resolver.pl?urn=nbn:de:bvb:19-epub-9702-3}
  {\enquote {\bibinfo {title} {Personenstand der
  {L}udwig--{M}aximilians--{U}niversit\"at {M}\"unchen. {W}inter--{H}albjahr
  1928/29. {A}. {B}eh\"ordenverzeichnis nach dem {S}tande vom 31. {D}ezember
  1928. {B}. {S}tudentenverzeichnis nach dem {S}tande vom 30. {N}ovember
  1928.}}\ }\bibinfo {note} {In German}\BibitemShut {NoStop}%
\bibitem [{\citenamefont {{LMU M\"unchen}}(1929)}]{LMU1929}%
  \BibitemOpen
  \bibfield  {author} {\bibinfo {author} {\bibnamefont {{LMU M\"unchen}},}}
  (\bibinfo {year} {1929}),\ \href
  {http://nbn-resolving.de/urn/resolver.pl?urn=nbn:de:bvb:19-epub-9703-8}
  {\enquote {\bibinfo {title} {Personenstand der
  {L}udwig--{M}aximilians--{U}niversit\"at {M}\"unchen. {S}ommer--{H}albjahr
  1929. {A}. {B}eh\"ordenverzeichnis nach dem {S}tande vom 1. {J}uli 1929. {B}.
  {S}tudentenverzeichnis nach dem {S}tande vom 31. {M}ai 1929.}}\ }\bibinfo
  {note} {In German}\BibitemShut {NoStop}%
\bibitem [{\citenamefont {Meitner}\ and\ \citenamefont
  {Hupfeld}(1931)}]{MeitnerHupfeld1931}%
  \BibitemOpen
  \bibfield  {author} {\bibinfo {author} {\bibnamefont {Meitner}, \bibfnamefont
  {L}}, \ and\ \bibinfo {author} {\bibfnamefont {H.~H.}\ \bibnamefont
  {Hupfeld}}} (\bibinfo {year} {1931}),\ \bibfield  {title} {\enquote {\bibinfo
  {title} {{\"U}ber das {S}treugesetz kurzwelliger $\gamma$--{S}trahlen},}\
  }\href {\doibase 10.1007/BF01520525} {\bibfield  {journal} {\bibinfo
  {journal} {Naturwissenschaften}\ }\textbf {\bibinfo {volume} {19}}~(\bibinfo
  {number} {37}),\ \bibinfo {pages} {775--776}},\ \bibinfo {note} {in
  German}\BibitemShut {NoStop}%
\bibitem [{\citenamefont {Oppenheimer}\ and\ \citenamefont
  {Plesset}(1933)}]{OppenheimerPlesset1933}%
  \BibitemOpen
  \bibfield  {author} {\bibinfo {author} {\bibnamefont {Oppenheimer},
  \bibfnamefont {J~R}}, \ and\ \bibinfo {author} {\bibfnamefont {M.~S.}\
  \bibnamefont {Plesset}}} (\bibinfo {year} {1933}),\ \bibfield  {title}
  {\enquote {\bibinfo {title} {On the {P}roduction of the {P}ositive
  {E}lectron},}\ }\href {\doibase 10.1103/PhysRev.44.53.2} {\bibfield
  {journal} {\bibinfo  {journal} {Phys. Rev.}\ }\textbf {\bibinfo {volume}
  {44}}~(\bibinfo {number} {1}),\ \bibinfo {pages} {53--55}}\BibitemShut
  {NoStop}%
\bibitem [{\citenamefont {Roqu\'{e}}(1997)}]{Roque1997}%
  \BibitemOpen
  \bibfield  {author} {\bibinfo {author} {\bibnamefont {Roqu\'{e}},
  \bibfnamefont {X}}} (\bibinfo {year} {1997}),\ \bibfield  {title} {\enquote
  {\bibinfo {title} {The manufacture of the positron},}\ }\href {\doibase
  10.1016/S1355-21989600021-4} {\bibfield  {journal} {\bibinfo  {journal}
  {Stud. Hist. Philos. M. P.}\ }\textbf {\bibinfo {volume} {28}}~(\bibinfo
  {number} {1}),\ \bibinfo {pages} {73--129}}\BibitemShut {NoStop}%
\bibitem [{\citenamefont {Thibaud}(1933)}]{Thibaud1933d}%
  \BibitemOpen
  \bibfield  {author} {\bibinfo {author} {\bibnamefont {Thibaud}, \bibfnamefont
  {J}}} (\bibinfo {year} {1933}),\ \bibfield  {title} {\enquote {\bibinfo
  {title} {L'annihilation des positrons au contact de la mati\`{e}re et la
  radiation qui en r\'{e}sulte},}\ }\href@noop {} {\bibfield  {journal}
  {\bibinfo  {journal} {C. R. Hebd. Acad. Sci.}\ }\textbf {\bibinfo {volume}
  {197}}~(\bibinfo {number} {25}),\ \bibinfo {pages} {1629--1632}},\ \bibinfo
  {note} {in French}\BibitemShut {NoStop}%
\bibitem [{\citenamefont {Thibaud}(1934{\natexlab{a}})}]{Thibaud1934b}%
  \BibitemOpen
  \bibfield  {author} {\bibinfo {author} {\bibnamefont {Thibaud}, \bibfnamefont
  {J}}} (\bibinfo {year} {1934}{\natexlab{a}}),\ \bibfield  {title} {\enquote
  {\bibinfo {title} {Positive {E}lectrons: {F}ocussing of {B}eams,
  {M}easurement of {C}harge--to--{M}ass {R}atio, {S}tudy of {A}bsorption and
  {C}onversion into {L}ight},}\ }\href {\doibase 10.1103/PhysRev.45.781}
  {\bibfield  {journal} {\bibinfo  {journal} {Phys. Rev.}\ }\textbf {\bibinfo
  {volume} {45}}~(\bibinfo {number} {11}),\ \bibinfo {pages}
  {781--787}}\BibitemShut {NoStop}%
\bibitem [{\citenamefont {Thibaud}(1934{\natexlab{b}})}]{Thibaud1934a}%
  \BibitemOpen
  \bibfield  {author} {\bibinfo {author} {\bibnamefont {Thibaud}, \bibfnamefont
  {J}}} (\bibinfo {year} {1934}{\natexlab{b}}),\ \bibfield  {title} {\enquote
  {\bibinfo {title} {Sur la d\'{e}mat\'{e}ralisation des \'{e}lectrons
  positifs},}\ }\href@noop {} {\bibfield  {journal} {\bibinfo  {journal} {C. R.
  Hebd. Acad. Sci.}\ }\textbf {\bibinfo {volume} {198}}~(\bibinfo {number}
  {6}),\ \bibinfo {pages} {562--564}},\ \bibinfo {note} {in French}\BibitemShut
  {NoStop}%
\bibitem [{\citenamefont {{Verband Deutscher Physikalischer
  Gesellschaften}}(1951)}]{DPG1951}%
  \BibitemOpen
  \bibfield  {author} {\bibinfo {author} {\bibnamefont {{Verband Deutscher
  Physikalischer Gesellschaften}},}} (\bibinfo {year} {1951}),\ \bibfield
  {title} {\enquote {\bibinfo {title} {1948 {B}lackett},}\ }\href {\doibase
  10.1002/phbl.19510071206} {\bibfield  {journal} {\bibinfo  {journal} {Phys.
  Bl.}\ }\textbf {\bibinfo {volume} {7}}~(\bibinfo {number} {12}),\ \bibinfo
  {pages} {560}}\BibitemShut {NoStop}%
\bibitem [{\citenamefont {Williams}(1934)}]{Williams1934}%
  \BibitemOpen
  \bibfield  {author} {\bibinfo {author} {\bibnamefont {Williams},
  \bibfnamefont {E~J}}} (\bibinfo {year} {1934}),\ \bibfield  {title} {\enquote
  {\bibinfo {title} {Scattering of {H}ard {G}amma {R}ays by {L}ead, and the
  {A}nnihilation of {P}ositive {E}lectrons},}\ }\href {\doibase
  10.1038/133415a0} {\bibfield  {journal} {\bibinfo  {journal} {Nature}\
  }\textbf {\bibinfo {volume} {133}}~(\bibinfo {number} {3359}),\ \bibinfo
  {pages} {415}}\BibitemShut {NoStop}%
\end{thebibliography}%

\end{document}